\documentclass[article,onecolumn]{revtex4}
\usepackage{graphicx}
\newcommand\beq{\begin{equation}}
\newcommand\bear{\begin{eqnarray}}
\newcommand\eeq{\end{equation}}
\newcommand\eear{\end{eqnarray}}
\begin{document}

\baselineskip=24pt

\title {Negative differential resistance in nanoscale transport in the Coulomb blockade regime}       

\author{\bf Prakash Parida, S Lakshmi\footnote
{ Presently, at Institute for Materials Science and Max
Bergmann Center of Biomaterials,  
\\Dresden University of Technology, D-01062 Dresden, Germany} and Swapan K Pati}

%\vspace*{0.1cm}

\address{Theoretical Sciences Unit and DST Unit on Nanoscience \\
Jawaharlal Nehru Centre For Advanced Scientific Research \\
Jakkur Campus, Bangalore 560064, India \\
E-mail: pati@jncasr.ac.in}

\date{\today}

\widetext

\begin{abstract}
\parbox{6in}

{\bf Abstract}

{Motivated by recent experiments, we have studied transport behavior of coupled
quantum dot systems in the Coulomb blockade regime using the master
(rate) equation approach. We explore how electron-electron interactions in a donor-acceptor 
system, resembling weakly coupled quantum dots with varying charging energy,
can modify the system's response to an external bias, taking it
from normal Coulomb blockade behavior to negative differential resistance (NDR)
in the curent-voltage characteristics.}
\end{abstract}
\maketitle

\narrowtext

The switching and negative differential resistance (NDR) behavior of nanoscale
systems has gained a lot of interest in the last decade, owing to the
potential applications in single molecule electronics and has been observed in a
variety of experimental systems, especially in the widely
studied Tour molecules \cite{Tour,Don}. There have been many theoretical studies to 
understand this phenomenon within mainly through the one-electron
picture \cite{Seminario,Cornil,rpati,lakshmi_PRB,lakshmi_jpc}. There have also been a number of 
theoretical studies on donor-acceptor double quantum dot systems, where strong 
rectification has been observed \cite{Cuniberti}, and others which showed NDR 
with variation in the dot-electrode coupling \cite{Thielmann, Aghassi,Paulsson} 
or due to a detuning of the dot levels \cite{Aghassi}. Another recent study has 
attempted to establish the conditions obeyed by the parameters involved, to find
such a collapse in the current magnitude\cite{Bhaskaran_generic}. 
Some recent experiments on double quantum dots also showed an NDR feature  \cite{Ono,tarucha} and
has rekindled interest in the phenomenon, occuring in the low temperature, 
weak-coupling limit. Theoretical studies of NDR in this single electron charging limit, is now  
gaining prominence and attracting a lot of research \cite{Hettler1,wegewis,Hung}. 
This regime, where mean-field descriptions usually fail, is one where electron 
charging  energies are very high as compared to the broadenings due
to average coupling, and are particularly important for small 
molecules which behave more like a quantum dot than a wire \cite{Datta_MB,Bhaskaran}.
Since mean field methods combined with standard non-equilibrium
Greens function (NEGF) \cite{negf1,negf3,negf4,negf5} treatment of 
transport is perturbative in the interaction parameter, it cannot capture the transitions between 
the spectrum of neutral and excited states, which can lead to a variety of interesting
features in the current-voltage characteristics. The formalism that has now come to be used widely to 
capture molecular transport in the Coulomb blockade regime is the master or rate equation method \cite{IEEE}.

\begin{figure}
\centering
\includegraphics[scale=0.6,angle=0.0]{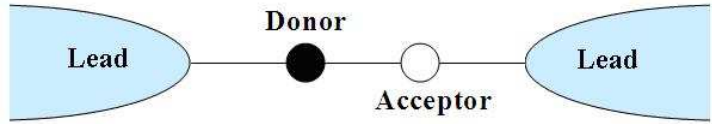}
\caption{\label{composite} A schematic representation of a two-dot system consisting of a donor and an acceptor 
coupled to two electrodes.} 
\end{figure}

 In this article, we use the above formalism to study a two-dot
system consisting of a donor and an acceptor (see the schematic given in figure 1) in the Coulomb blockade regime.
Taking cue from our previous mean-field transport studies on two-level systems
which showed interesting non-linear behavior in their current-voltage
characteristics \cite{lakshmi_PRB}, here we explore the role of strong correlations in
affecting their transport behavior. This study becomes interesting, especially in the 
context of the difference in
their low-lying excitations, which would play a very important role in their
low-bias current-volatage characteristics. The rate equation formalism describes 
transport through a correlated system with many-body eigenstates. The presence of Coulomb 
interactions results in occupation probabilities of each many body state that 
cannot be factorized as the product of the occupation probabilities of each single 
electron level. Hence, in this case, the full rate-equation problem, where the 
occupation probability of each many-body state is treated as an
independent variable is solved, neglecting off-diagonal coherences.
In this method, the transition rate, $\Sigma_{{s^\prime}\rightarrow{s}}$,
from the many-body state $s^\prime$ to $s$, differing by one electron, is calculated up to linear order
in $\Gamma$ (which is the bare electron tunneling rate between the system and the electrodes), 
using Fermi's golden rule as \cite{Beenakker},

\begin{eqnarray}
\Sigma_{{s^\prime}\rightarrow{s}}^{L+}=\Gamma f_L(E_s-E_s^\prime) \sum_{\sigma} 
|<s|C^\dag_{1\sigma}|s^\prime>|^2\nonumber  
%\eear
\end{eqnarray}
\begin{eqnarray}
%\bear
\Sigma_{{s^\prime}\rightarrow{s}}^{R+}=\Gamma f_R(E_s-E_s^\prime) \sum_{\sigma} 
|<s|C^\dag_{N\sigma}|s^\prime>|^2 
\end{eqnarray}

\noindent with a corresponding equation for
$\Sigma_{{s} \rightarrow {s^\prime}}^{L-}$ and
$\Sigma_{{s} \rightarrow {s^\prime}}^{R-}$ obtained
by replacing $f_{L,R}(E_s-E_s^\prime)$ by
$(1-f_{L,R}(E_s-E_s^\prime))$. Here, $+/-$ correspond to the
creation/annihilation of an electron
inside the dot due to electron movement from/to left (L) or right (R) electrode.
$C^\dag_{1\sigma}$ and $C^\dag_{N\sigma}$ are the creation operators for electrons of spin, $\sigma$
at the first and Nth sites respectively. 
We have also assumed that the creation and annihilation happen only at the 
terminal sites.
The total transition rate is then obtained as,
$\Sigma_{{s} \rightarrow {s^\prime}}=\Sigma_{{s} \rightarrow {s^\prime}}^{L+}+
\Sigma_{{s} \rightarrow {s^\prime}}^{R+}+\Sigma_{{s} \rightarrow {s^\prime}}^{L-}+
\Sigma_{{s} \rightarrow {s^\prime}}^{R-}$.
The non-equilibrium probability $P_s$ of occurrence of each many-body state $s$ is obtained
by solving the set of independent rate equations defined by
$\dot{P_s}=\sum_{s^\prime}(\Sigma_{{s^\prime} \rightarrow s}P_{s^\prime}-\Sigma_{{s} 
\rightarrow {s^\prime}} P_s)$
\noindent through the stationarity condition $\dot{P_s}=0$ at steady state. 
This results in a homogeneous set of equations of the size of the many-body space. Taking advantage
of the normalization condition $\sum_s{P_s}=1$, we obtain linear equations,
which can be solved using well-known linear algebraic methods.
The steady state probabilities are then used to obtain the terminal current as,

\begin{eqnarray}
I_{\alpha}={e \over \hbar} \sum_{s,s^\prime}\Sigma_{{s^\prime} \rightarrow s}^{\alpha+}
P_{s^\prime}-\Sigma_{{s} \rightarrow {s^\prime}}^{\alpha-} P_s
\end{eqnarray}

\noindent
where $\alpha=L/R$. Using the above prescribed method, we study a two site system
described by the Hamiltonian,
\begin{eqnarray}
H =  \sum_{i=1}^2\ (\epsilon_i-eW_g)a^{\dag}_{i}a_{i}+ \sum_{\sigma=\uparrow, \downarrow} -
t(a^{\dag}_{1\sigma}a_{2\sigma}+h.c.)
 + U \sum_{i=1}^2 n_{i\uparrow}n_{i\downarrow}+ V_{12}(n_1-\bar{n})(n_2-\bar{n}) 
\end{eqnarray}
\noindent
where $t$ is the hopping strength between the sites with same spin ($\sigma$), $\epsilon_{1,2}$
are the on-site energies, $U$ is the Hubbard interaction between
electrons at the same site, $V_{12}$ is the nearest-neighbor Coulomb interaction and $W_g$ is the external gate bias. The
average charge $(\bar{n})$ is assumed to be unity here \cite{dotcharge}. $\bar{n}$ actually gives a constant shift to 
the energy levels with fixed number of electrons. For two sites with two electrons, the energy levels are negatively shifted by 
$V_{12}$ amount. Note that, there exists two 
quantum phases in this model with variation of interaction parameters. For the half filled ground state, 
with zero onsite energies, while $U > V_{12}$/2 represents a spin density wave (SDW) phase, 
$U < V_{12}$/2 corresponds to charge density (CDW) phase in thermodynamic limit\cite{Hirsch,Tsuchiizu}. However, in
our case with two sites, while for $U < V_{12}$, the half filled ground state gives higher preference to the state with two 
electrons of opposite spins at one site, for $U > V_{12}$, the state with one electron at each site is more prefered.    

        To study the transport properties through a double quantum dot system comprising of a donor and an acceptor in the 
weak coupling regime, we parameterize the different coupling strengths in the total system (system+leads). For perturbation theory 
to be valid at temperature T, we ensure that $\Gamma \ll k_BT $. More specifically in our calculations,  
we use the value of $\Gamma=0.25meV$ for T=300K and $\Gamma=0.01meV$ for T=0.66K, which are also much smaller than the corresponding
charging energies, e.g. Hubbard $U$. As our primary interest focuses on NDR effect in the system, we choose   
asymmetry in the onsite energy ($\Delta \epsilon$ =$\epsilon_2$ -$\epsilon_1)$ to be larger than the 
interdot hopping parameter (t), and  vary the Hubbard $U$ around $\Delta \epsilon$.

\begin{figure}
\centering
\includegraphics[scale=0.60,angle=0] {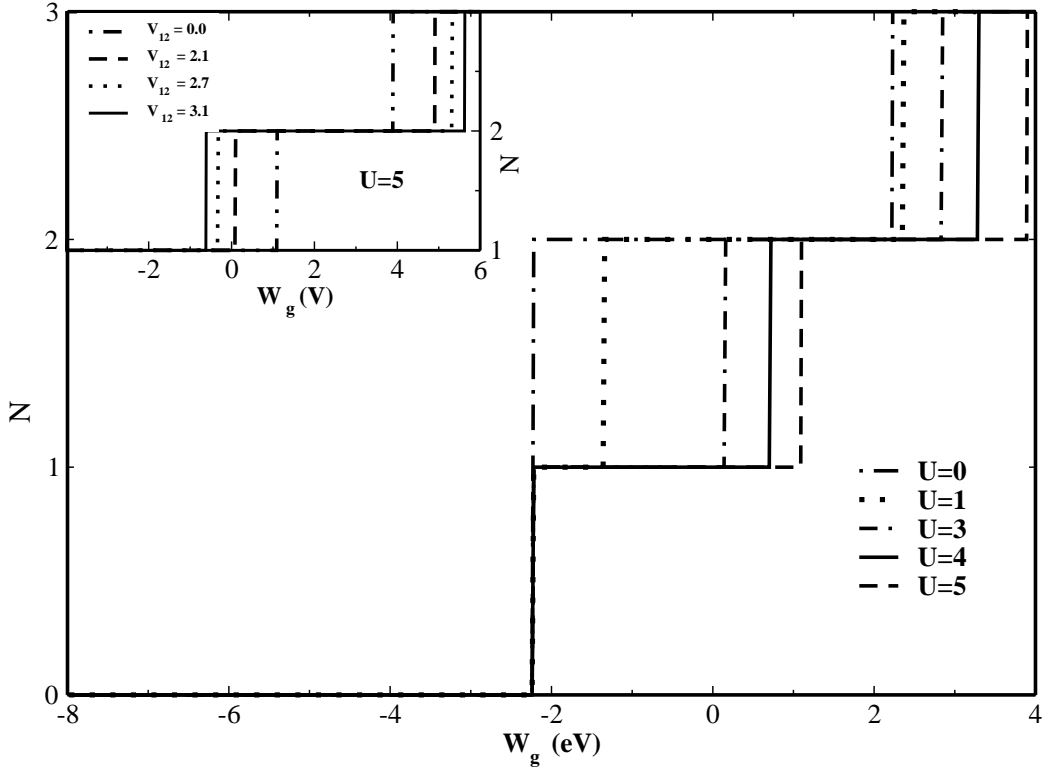}
\caption{ \label{number_vg}  The number of electrons (N) in a 2 site donor-acceptor system with
variation of gate bias of the system ($W_g$) for various values of the Hubbard prameter $U$.   
The inset shows the same for $U$ = 5 eV with different values of nearest neighbor Coulomb interaction parameter, $V_{12}$.  
Here, $\epsilon_2=-\epsilon_1= 2.0$ eV, $t=1.0$ eV and all the values of $U$ and $V_{12}$ are in eV ($\bar{n}=1$).} 
\end{figure}

\begin{figure}
\centering
\includegraphics[scale=0.6,angle=0]{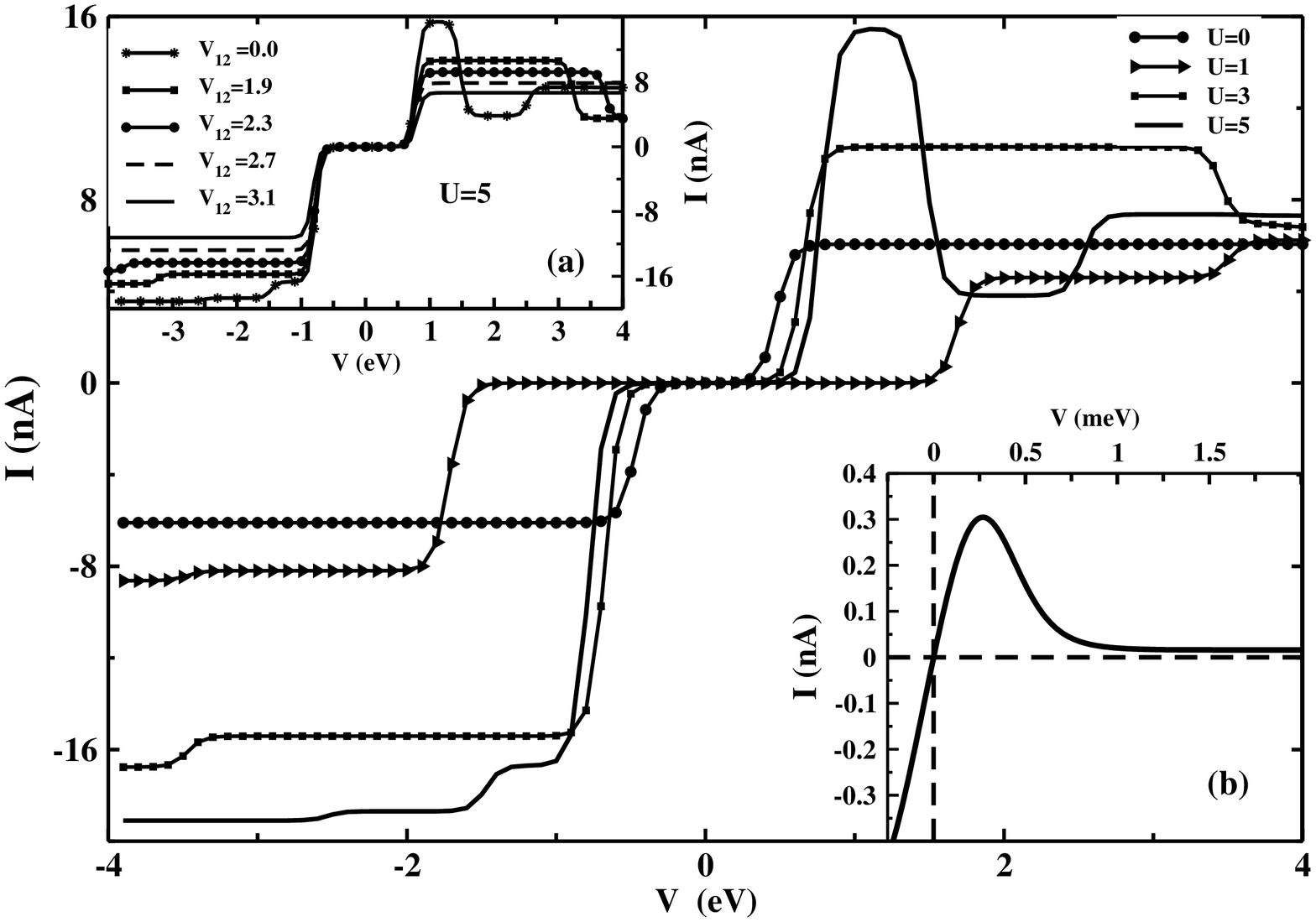}
\caption{ \label{iv_uv} Current (I) - source-drain voltage (V) characteristics of the 2 site donor-acceptor
system for various values of $U$ and the inset (a) represents the same for $U$ = 5 eV with various $V_{12}$ values, for 
$\epsilon_2=-\epsilon_1=2.0eV, t=1eV, \Gamma=0.25 meV$ and Temperature (T)=300K.   
The inset (b) shows the same, for $\epsilon_2=-\epsilon_1=2meV, t=0.2meV, U=4meV,  \Gamma=0.01 meV$ and Temperature (T)= 0.66K}.
\end{figure}

     We  adopt the well known exact diagonalization (ED) method to solve the Hamiltonian in equation 3 
for the system containing two sites. As the total number of electrons, N, 
and z-component of the total spin, $S_z$, commute with the Hamiltonian (H) and can be considered as 
conserved quantities, the H matrix can be diagonalized for a particular charge and spin sector. 
The Fock space 
then can be factored into many blocks, with largest block consisting of 4 states 
with quantum numbers, number of electron (N)=2 and $S_z$ =0. The ground state energies for 
N=1 ($E_{1e}$), N=2 ($E_{2e}$) and N=3 ($E_{3e}$) with 
onsite energies $\epsilon_1= \epsilon_2$ =0 can be easily found ($\bar {n}=1$)

\begin{eqnarray}
E_{1e} = -t-W_g \nonumber \\
E_{2e} = \frac{U-V_{12}}{2} - \sqrt{\frac{{(U-V_{12})}^2}{4} + 4t^2} - 2W_g\nonumber \\
E_{3e} = U-t- 3W_g
\end{eqnarray}

\noindent Thus, the gate bias 
window ($\Delta W_g= E_{3e}^0 + E_{1e}^0-2E_{2e}^0$) over which the N=2 (half-filled) state becomes lowest energy state 
can be estimated to be,
\begin{eqnarray}
\Delta W_g = -2t + V_{12} + \sqrt{{(U-V_{12})}^2 + 16t^2}
\end{eqnarray}
\noindent which strongly depends on the parameters involved. $ E_{1e}^0, E_{2e}^0$ and $ E_{3e}^0$ are the 
ground state energies for 1e, 2e and 3e states respectivily in the absence of gate bias.
However, with inclusion of asymmetric on-site energies, 
the general analytical expression for the energies become quite lengthy. For a chosen onsite energy values 
considering donor and acceptor sites, $\epsilon_2=-\epsilon_1= 2.0eV$, we plot in figure \ref{number_vg} 
the number of electrons in the lowest energy state as a function of gate bias. This is obtained by calculating 
the many-body states with minimum energy at every value of gate bias $W_{g}$ as Min ($E_{s}$). 
For $U<4eV$, with increase in onsite electron-electron interaction, there is a reduction 
in gate bias over which two electron state is stable, while on the contrary, for $U>4eV$, the gate bias window increases with increase in 
onsite electron-electron interaction. Furthermore, for $V_{12}$ values closer or greater than $U$/2, 
the bias range over which two electron state is the lowest energy state, increases with increase in $V_{12}$.
This happens, because an increases in $U$ by 1eV causes no change to $E_{1e}^0$, an increase of 1eV to $E_{3e}^0$, 
but an increases of more than 0.5eV to $E_{2e}^0$ for $U<4eV$ and an increase of less than 0.5eV to $E_{2e}^0$ 
for $U>4eV$. This is due to the fact that, for $U<4eV$, two electron ground state gives higher preference to the state with 
two electrons of opposite spins at the site with lower on-site energy. However, for $U>4eV$, the it prefers the state  
with one electron each at the donor and at the acceptor. 
Hence for $U<4eV$, an increase in $U$ value by 1eV, causes an increase in the value of 2$E_{2e}^0$ by more than 1eV and 
an increase of 1eV to $E_{3e}^0$, so effectively reducing the value of $\Delta W_g$.
However for  $U>4eV$, the increase in the value of 2$E_{2e}^0$ is always less than 1eV and hence $\Delta W_g$ increases with 
increase in $U$ values.

\begin{figure}
\centering
\includegraphics[scale=0.6,angle=0.0]{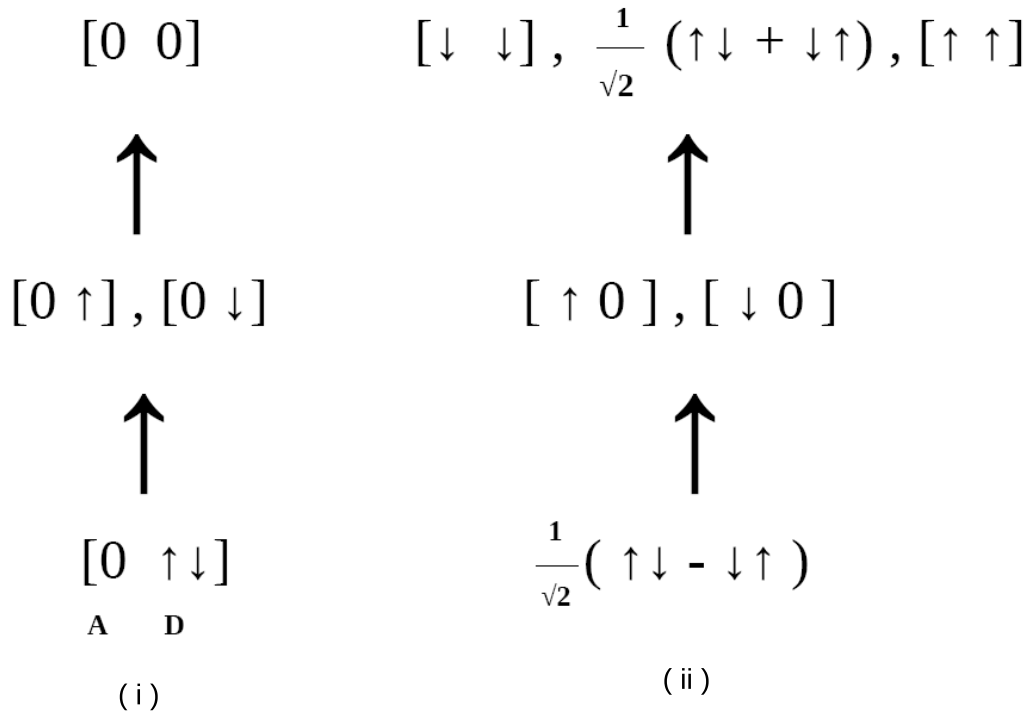}
\caption{A schematic describing the transitions between the states of the donor-acceptor
system, (i) in the small $U$ regimes (ii) in large $U$ regimes. The arrow direction indicates
the states to which the transition occurs.}
\end{figure}

      For obtaining current, for every value of $U$ and $V_{12}$, the
Fermi energy ($E_{F}$) is chosen as the value of the gate bias which ensures that two electron state 
is the ground state. The Fermi energy is also placed  in such a way that we observe transition
from the ground state to the state with one less electron. After fixing the Fermi energy, we have studied the 
current as a response of source-drain bias (V) in  all our calculations.  
In figure \ref{iv_uv}, we have plotted the
I-V characteristics of the system for a range of $U$ and $V_{12}$ values at room temperature.
As can be seen clearly, low values of $U$ results in step-like features in 
I-V characterestics, while with increase in $U$, a rise
and fall in current (a NDR feature) is observed for positive values of source-drain bias.
Interestingly, with inclusion of nearest-neighbor Coulomb interaction, $V_{12}$,  
the I-V characteristics show wide plateau region before showing NDR feature. 
However, the height of the NDR peak decreases in the positive source-drain bias region with increase in $V_{12}$.
To compare our results with the experimental findings at low temperature, in the inset(b) of figure \ref{iv_uv}, 
we  have plotted the low temperature behaviour of I-V characteristics. Note that, the NDR peak 
together with the overall I-V feature compare fairely well with the experimental results obtained by Tarucha et al
on GaAs-based double quantum dots \cite{tarucha} . 
We also note that, there is in fact no qualitative change in the I-V characteristics except for a constant shift in bias, 
if we change average dot charge($\bar {n}=1$) from one to zero in the Hamiltonian in equation 3.

\begin{figure}
\centering
\includegraphics[scale=0.6,angle=0.0]{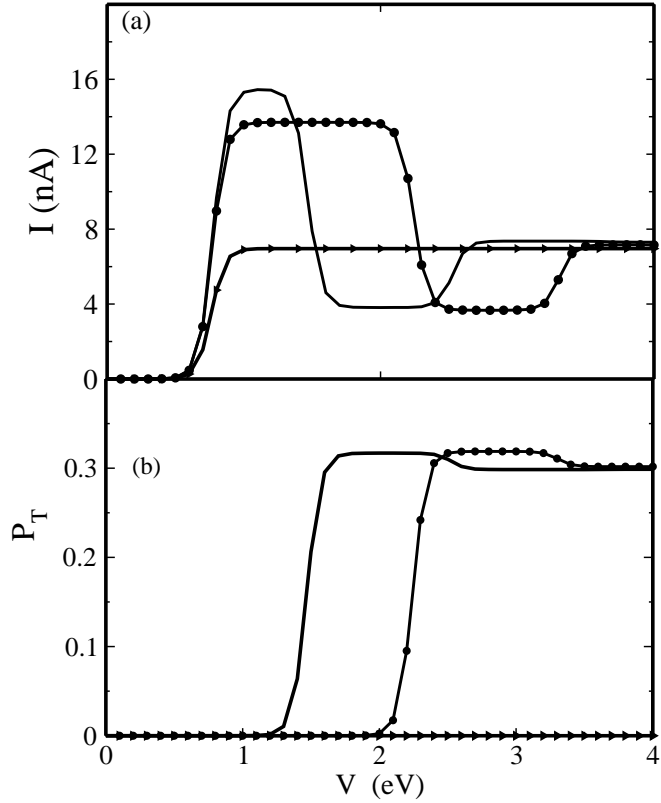}
\caption{\label{prob} The variation of (a) current (I) (b) occupation probability ($P_T$) of the 2$e$
triplet state with source-drain bias, corresponding to $U=5$eV and varying $V_{12}$ values: $V_{12}$=0 eV (solid line), 
$V_{12}$=1 eV (circle) and $V_{12}$=3 eV (triangle).}
\end{figure}

    The step like feature in I-V is well understood in literature, as due to Coulomb repulsions
\cite{Ratner,Datta2}. However, to understand the NDR feature, we analyze the probabilties of 
occurance of various many body states. We find that NDR occurs when the source-drain bias drives the system 
from the 1$e$ doublet to a higher excitation of the 2$e$ state, namely 
the triplet states, instead of to the state with zero electron. It is because, when $U$ is small,
the ground state gives higher preference to the state with two electrons of opposite spins at 
the site with lower on-site energy. This allows annihilation of an electron by the electrode 
followed by one more annihilation leading to a transition from the 2$e$ singlet to 
the 1$e$ doublet and then to the state with zero electron. However, when $U$ increases, the ground state
gives more weightage to the state with one electron at the donor and one at the acceptor. 
This allows for one electron annihilation from the ground state to the
1$e$ doublet state, followed by a creation of an electron from the same electrode
to the 2$e$ triplet state, which has the same energy as the zero electron state.
Since the current at any electrode is calculated at steady state as the
difference between the outgoing and incoming current, this transition results in a reduction 
in current leading to the negative differential resistance (NDR) peak in large $U$ limit. 
A schematic figure 
describing the states involved with increase in  positive source-drain bias for small and large $U$ limits
are shown in figure 4.
Note that, with inclusion of $V_{12}$, particularly for large $V_{12}$ values, the charge density modulated state
gets prominance, similar to the ground state electronic configuration as in the small $U$ limit. 
Thus, with increase in $V_{12}$, the NDR feature gets suppressed. Also, since with inclusion of $V_{12}$, the gate 
bias range over which the 2$e$ state remains the ground state differs, we pin the electrode's Fermi energy 
in such a way that the transition from 2$e$ singlet to 1$e$ doublet state occurs at the same values 
of V (see the inset (a) of figure 3)
for a range of $V_{12}$ values. However, with increase in positive bias, the electrochemical potential 
at the left 
electrode ($\mu_L$) moves down and that at right electrode ($\mu_R$)
moves up, causing the transport channel $\epsilon=E_T-E_{1e}$ 
is in resonance with the levels of the electrodes, where $E_T$ and $E_{1e}$ are 
the energy levels associated with 2$e$ triplet state and 1$e$ doublet state respectively. 
With large $V_{12}$ values, this channel width causes the plateau in I-V characterestics to be 
wider before showing NDR.

    To understand the NDR feature more clearly, and to estimate the height of the peak value in the I-V
plot in figure 3, we calculate the probability of occurance of the 2$e$ triplet state with increase in V
for a range of Hamiltonian parameters. In figure \ref{prob}, we have plotted the variation of current and 
the occupation probability of 2$e$ triplet state with source-drain bias for $U$ = 5 eV. It is clear that, when 
the occupation probability of the 2$e$ triplet state starts increasing appreciably, the current decreases 
in magnitude, however, only to a nonzero value. The main point is that the triplet state being the 
blocking state suppresses the current, however, since its probability of occurance does not increase 
more than 30\%, there is still some finite current (leakage current) 
which flows through the system. Note that, the I-V characteristics are asymmetric because of the inherent asymmetry in the system 
comprising of a donor and an acceptor with different site energy ($\epsilon$) values.

     In conclusion, we have studied transport behavior of donor-acceptor system in the
Coulomb blockade regime through the rate equation approach.
Our study shows how the variation in the on-site Coulomb repulsions can influence the system's
response to an external source-drain bias. A strong Coulombic repulsion
even results in NDR for positive values of source-drain bias in the I-V characteristics.
Also, a strong nearest-neighbour Coulomb interaction suppresses the
NDR like feature, taking back the system to normal Coulomb staircase regime.

Acknowledgement: PP acknowledges the CSIR for a research fellowship and
SKP acknowledges research support from CSIR and DST, Government of India and AOARD, Asian Office.

\end{document}